\documentclass[journal,onecolumn]{IEEEtran}
\usepackage{booktabs}
\usepackage{subfigure}
\usepackage{amsmath}                
\usepackage{amsthm}                 
\usepackage{amssymb}
\usepackage{authblk}
\usepackage{thmtools}
\usepackage{kpfonts}
\usepackage[geometry]{ifsym}
\usepackage{dsfont}
\usepackage{multirow}
\usepackage[mathscr]{euscript}
\usepackage{graphicx}
\usepackage{color}
\usepackage[inline]{enumitem}
\usepackage{framed}
\usepackage{float}
\usepackage{longtable}
\usepackage{cite}
\usepackage{enumitem}
\usepackage{caption} 
\usepackage{circuitikz}
\usepackage{tikz}

\usetikzlibrary{patterns,decorations.pathmorphing,positioning, decorations.markings}
\usepackage{xcolor}

\usepackage[colorlinks=true, citecolor=blue, linkcolor=blue]{hyperref}       
\usepackage[font=itshape]{quoting}
\usepackage{lscape}
\usepackage{makecell}
\usepackage[section]{placeins}

\usepackage{listings}
\usepackage{comment}
\usepackage{framed} 
\usepackage{nomencl} 
\makenomenclature
\allowdisplaybreaks
\declaretheoremstyle[%
  headfont=\bfseries,%
  headpunct={:},%
  notefont=\normalfont\bfseries,%
  notebraces={--~}{},
    qed=$\blacksquare$,
]{definitionstyle}
\theoremstyle{definition}
\declaretheorem[style=definitionstyle,name=Definition]{defn}

\theoremstyle{definition}

\theoremstyle{plain}
\theoremstyle{remark}

\newcommand{\liinesbigfig}[4]{\begin{figure*}[ht!]\begin{center}\includegraphics[width=#4in]{#1}\vspace{-0.1in}\caption{#2}\label{#3}\end{center}\vspace{-0.2in}\end{figure*}}

\begin{document}
%
\title{Embedding Economic Input–Output Models in Systems of Systems: An MBSE and Hetero-functional Graph Theory Approach}

\author[1]{Mohammad Mahdi Naderi\textsuperscript{*}}
\author[1]{Megan S. Harris}
\author[1]{John C. Little}
\author[2]{Amro M. Farid}

\affil[1]{Department of Civil and Environmental Engineering, Virginia Tech, Blacksburg, Virginia 24061, United States}
\affil[2]{School of Systems and Enterprises, Stevens Institute of Technology, Hoboken, New Jersey, 07030, United States}


\date{September, 2024}
\maketitle

\begin{abstract}
Characterizing the interdependent nature of Anthropocene systems of systems is fundamental to making informed decisions to address challenges across complex ecological, environmental and coupled human-natural systems. This paper presents the first application of Model-Based Systems Engineering (MBSE) and Hetero-functional Graph Theory (HFGT) to economic systems, establishing a scalable and extensible methodology for integrating economic input–output (EIO) models within a unified system-of-systems modeling framework. Integrating EIO models into the MBSE–HFGT workflow demonstrates how the structural form and function of economic systems can be expressed through SysML’s graphical ontology and subsequently translated into the computational structure of HFGT. Using a synthetic Rectangular Choice of Technology (RCOT) example as a pedagogical foundation, the study confirms that the dynamics captured by basic EIO models, as well as other complex economic models grounded in EIO theory, can be equivalently reproduced within the MBSE-HFGT framework. The integration with MBSE and HFGT thus preserves analytical precision while offering enhanced graphical clarity and system-level insight through a shared ontological structure. By integrating modeling languages and mathematical frameworks, the proposed methodology establishes a foundation for knowledge co-production and integrated decision-making to address the multifaceted sustainability challenges associated with Anthropocene systems of systems.
 
\end{abstract}

\section{Introduction}
Humans are the primary drivers of profound changes in Earth's systems, giving rise to myriad interconnected societal challenges that characterize the Anthropocene \cite{little2023earth,verburg2016methods,steffen2018trajectories}. The concept of the Anthropocene describes the current stage of Earth system change in which human activities have emerged as a dominant driving force \cite{zalasiewicz2020anthropocene,liu2007coupled,steffen2007anthropocene,goudie2018human,lewis2018human}, influencing geophysical, biophysical, socioeconomic, sociocultural, and socio-technical processes \cite{little2023earth}. Rapid population growth, widespread urbanization, and intensified resource exploitation are among human-induced pressures \cite{mondal2022challenges} that reshape the Earth's system trajectories and exacerbate intertwined social, ecological, and technological challenges \cite{kotze2019earth,mazac2020post}.

These challenges are inherently interconnected, as disturbances in one system can cascade across the entire system of systems (SoS), generating complex feedback and interactions \cite{folke2021our,williams2015anthropocene,malhi2017concept}. For example, rapid economic growth contributes to climate change by requiring mass production, destabilizing local hydrological cycles, and changing energy-use regimes \cite{fankhauser2005climate,nasir2019role}. The spectrum of Anthropocene challenges is extensive, encompassing global warming, overexploitation of natural resources, freshwater scarcity, habitat degradation, and widespread environmental pollution \cite{steffen2007anthropocene,toivanen2017many}. These pressures threaten numerous endangered species and compromise human well-being \cite{steffen2011anthropocene}. Nonetheless, these challenges are frequently addressed in isolation, with relatively few studies adopting integrative approaches that explicitly account for their interdependencies \cite{little2023earth, an2014agent, steffen2011anthropocene, reyers2018social, assembly2015transforming}. Although recent research acknowledges the interconnected nature of Anthropocene challenges \cite{xiaoming2018linking,li2023challenges,liu2007coupled}, capturing synergies and trade-offs across systems remains complicated due to limited understanding \cite{berkes2008navigating,robinson2018modelling} and the absence of analytical frameworks capable of quantifying these complex relationships \cite{little2023earth, li2023challenges}.

The economic system constitutes a foundational and inseparable element of the Anthropocene SoS \cite{liu2007complexity,leach2018equity}. It functions as a primary driver of global change while remaining intrinsically interwoven with and biophysically constrained by social and natural systems \cite{wironen2020critically, costanza1993modeling}. This relationship is bidirectional\cite{bocta2018bidirectional}, as the economic system exerts profound impacts on and is significantly influenced by the natural and social spheres, constituting a coupled social–economic–natural SoS\cite{wang2018structure,holling2001understanding}. One manifestation of this is economic growth in the Anthropocene, particularly fossil-fuel-driven growth, which is responsible for the deterioration of Earth systems and the transgression of planetary boundaries \cite{knight2014economic,abidoye2015climate}. For example, studies indicate that projected biodiversity loss exceeds safe thresholds globally and rises significantly with GDP per capita, suggesting that continued economic expansion beyond planetary boundaries undermines the planet’s carrying capacity \cite{sol2019economics}.

In contrast, the economic system is significantly influenced by the other systems through feedback mechanisms \cite{verburg2016methods,kellie2011emergent}. For example, resulting environmental changes—such as global warming and extreme events—can suppress economic growth by inducing climate-related damages that compromise water resources, food production, and human health \cite{stern2008economics,tol2018economic}. Successfully anticipating the consequences of these interactions between the economic system and other interlinked systems, and effectively addressing sustainability challenges, requires models that go beyond traditional approaches, which often treat anthropogenic drivers as exogenous to the economic system \cite{liu2007complexity,boumans2015multiscale,sherwood2020putting}. Consequently, there is a critical need to develop SoS models that explicitly capture nonlinear, complex dynamics and interdependencies among the economic system and other interconnected systems \cite{holling2001understanding}.

Numerous studies have examined unidirectional and bidirectional causal relationships between the economy and hydrological\cite{baker2021hydro,harou2009hydro,esteve2015hydro,heinz2007hydro}, energy\cite{nakata2004energy,del2018modelling,morris2019representing,gemelli2011gis}, food\cite{hamam2021circular}, and other interconnected systems\cite{chai2020quantifying, chen2022evaluation,sun2022new,akhtar2013integrated} to assess trade-offs and synergies.  However, only a limited body of research explicitly conceptualizes the economy as an integral component of the broader complex SoS in the Anthropocene, in which multidirectional causalities link economic dynamics to other interwoven systems. In most of these studies, this lack of explicit integration is reflected in prevailing modeling practices \cite{donges2018taxonomies,pan2025evolution}. This highlights a predominant reliance on exogenously defined drivers, constraining the ability to capture the full complexity of cross-system interdependencies. Such limitations arise because modeling approaches for coupling the economic system with other entangled systems often:
\begin{enumerate*}[label=(\roman*)]
    \item lack the capacity to represent the full detail of multiple interacting domains, including economy\cite{verburg2016methods},
    \item require simplification of key elements to reduce complexity\cite{amaya2021coupled}, and
    \item face technical and interoperability barriers, as linking distinct models typically demands specialized interfaces \cite{li2023challenges}.
\end{enumerate*}
Effective coupling strategies must balance complexity, functionality, and computational efficiency while simultaneously maximizing cross-disciplinary knowledge integration \cite{punjabi2025conceptual} through a shared common language \cite{yang2019ontology} and ensuring the internal consistency of the integrated SoS \cite{naderi2025convergent}.\\

\subsection{Objective and Original Contribution}

This paper presents the first application of Model-Based Systems Engineering (MBSE) and Hetero-functional Graph Theory (HFGT) to economic I-O (EIO) systems, and demonstrates that MBSE-HFGT is well-suited to fully internalize EIO models while capturing the heterogeneous dynamics among economic systems and other interwoven Anthropocene SoS. The paper develops a framework for embedding EIO models as key components of complex SoS in the Anthropocene, and demonstrates how these models can be systematically integrated into an MBSE–HFGT workflow. MBSE provides a formal, ontology-driven environment capable of capturing form, function and heterogeneous interactions of SoS, enabling all systems—including the economic system—to be represented within a shared common language. 

The workflow begins by graphically representing the EIO model in the Systems Modeling Language (SysML) to capture the form and function of the economic system. HFGT is then applied to translate these SysML models into a rigorous computational framework. While MBSE can be used as a standalone methodology, this paper shows how coupling MBSE with HFGT provides a powerful quantification tool that preserves the analytical insights of basic EIO models. The MBSE–HFGT framework, therefore, overcomes limitations identified in prior studies by ensuring ontological consistency between the economic system and other heterogeneous systems, preserving the analytical rigor of EIO models, and extending them to an SoS context as endogenous components, without requiring external interfaces or simplifying assumptions.

To illustrate this integration, the methodology is demonstrated using a synthetic EIO model commonly used in the literature by modelers across disciplines to evaluate sectoral economic dynamics. This example confirms that MBSE–HFGT not only can reproduce the analytical insights of economic modeling within an integrated SoS platform but also offers greater graphical clarity and system-level insight through a shared ontology that facilitates the co-production of knowledge and cross-disciplinary understanding. In doing so, the paper bridges a critical gap between economic modeling and systems engineering, establishing a unified ontological and mathematical foundation for integrated decision-making in complex socio-economic–environmental Anthropocene SoS.

\subsection{Paper Outline}
The remainder of the paper is organized as follows. Sec. \ref{Sec:EIOfundamentals} introduces the fundamentals of EIO models. Sec. \ref{Sec:MBSEConcept} elaborates on MBSE and graphical models of system form and function in SysML. Sec. \ref{Sec:HFGTConcept} presents essential background on HFGT, and Sec. \ref{Sec:Methodology} provides a synthetic, motivational RCOT example and demonstrates how MBSE-HFGT can be applied to internalize it. Sec. \ref{Sec:DiscussionandResults} discusses the implications of these results for simulating EIO models and for addressing societal challenges that cascade through interconnected systems. Sec. \ref{Sec:Conclusion} brings the paper to a conclusion.

\section{Conceptual and Theoretical Background}\label{Sec:TheoreticalBackground}

\subsection{Economic Input–Output (EIO) Model}
\label{Sec:EIOfundamentals}

EIO modeling, originally developed by Leontief (1936)\cite{leontief1936quantitative}, captures the interdependencies of production and consumption within an economic system\cite{miller2009input}. Over time, EIO models have been increasingly utilized in economic–environmental analyses to evaluate environmental sustainability\cite{mattila2010quantifying,liang2012comparisons,wiedmann2009companies}, energy management \cite{li2021multi, guevara2017multi}, risk assessment \cite{ma2012integrating,anderson2007risk}, and complex nexus systems \cite{wang2022multivariate,chen2023multi,wang2019energy}. EIO models evaluate direct, indirect, and induced impacts, which together provide a more comprehensive view of the dynamics of an economic system \cite{lenzen2003environmental, wiedmann2007examining}. The scientific literature identifies three main types of EIO models: basic (monetary) EIO, physical EIO, and hybrid EIO models \cite{miller2009input}. The basic EIO model accounts for intersectoral flows in monetary units (e.g., euros), the physical EIO model in physical units (e.g., tons), and the hybrid EIO model in a mixed-unit framework, combining monetary and physical units as needed\cite{liang2017structural,hoekstra2005economic,dietzenbacher2009physical,miller2009input}. These models share the common foundation of EIO accounting, which captures the balance between the supply and use of products across sectors. 

EIO models are based on the product supply–use equilibrium published in EIO tables. The equilibrium can be expressed as:  
\begin{equation}
\label{eq:leo1}
{x} = {Z} \mathds{1} + {y}
\end{equation}
where ${x}$ is a column vector of economic output with $n$ rows corresponding to industries, ${Z}$ is an $n \times n$ matrix of intermediate sales, ${y}$ is a column vector of final demand with $n$ rows, and $\mathds{1}$ is a column vector of ones.  

The technical coefficients matrix ${A}$ represents the share of goods and services required for production.  Its elements $a_{ij}=z_{ij}/x_{j}$ where $z_{ij}$ denotes the quantity of intermediate goods $i$ sold by industry $j$, and $x_j$ represents the total economic output of industry $j$. The model assumes that the matrix of technical coefficients is constant, reflecting the model's linearity.
Finally, Eq. \ref{eq:mainleo} is obtained by substituting the technical coefficients matrix $A$ into Eq. \ref{eq:leo1} and performing factorization.  
\begin{equation}
\label{eq:mainleo}
{x} = ({I} - {A})^{-1} {y} = {B}{y}
\end{equation}
where ${B} = ({I} - {A})^{-1}$ is the Leontief inverse matrix. Its elements $b_{ij}$ quantify the total, direct and indirect,  effect on the output of industry $j$ due to a one-unit change in final demand for product $i$.  Additionally, to compute the factor use vector $\boldsymbol{\phi}$, the final output vector ${x}$ is multiplied by the factor requirement matrix ${F}$:
\begin{equation}
\label{eq:factoruse}
\boldsymbol{\phi} = {F}{x}
\end{equation}

Building on this foundation, basic EIO modeling can be extended to economic input-output life-cycle assessment (EIO-LCA)\cite{egilmez2016fuzzy,shi2019economic}, multi-region input–output (MRIO)\cite{wiedmann2009review,lenzen2013building}, rectangular choice of technology (RCOT)\cite{duchin2011sectors,springer2018price}, environmentally-extended input-output (EEIO)\cite{chen2018global,kitzes2013introduction,yang2017useeio} and ecologically-extended input-output (ECEIO) \cite{liu2019integrated}, among others.

The RCOT model, as an extension of the basic EIO formulation, endows each sector with the capacity to select among multiple alternative technologies \cite{springer2018price, ju2019revealing, amaya2022applying}. To do this, the basic EIO model can be reformulated to accommodate alternative technologies by expanding the matrices I, A, and F with additional columns representing these technological options, thereby inducing a corresponding augmentation of the final output vector, x\cite{duchin2011sectors}.

To demonstrate this, each sector $i$ can choose from among $T$ alternative technologies. The resulting model can be expressed accordingly as a generalization of Eq. \ref{eq:mainleo}:
\begin{align}\label{firstconseio}
({\cal I^*} - A^*) x^* \ge y \\
f \le F^* x^*,
\label{secondconseio}
\end{align}
where $x^*$ is a $T \times 1$ vector, {$ \cal I^*$} and $A^*$ are $n \times t$ matrices, $y$ remains $n \times 1$, $F^*$ is a $k \times t$ matrix, and $f$ remains $k \times 1$.  In general, $k$ increases relative to the basic model if some technology-specific factors are included for the new options. 

The objective is formulated as the minimization of total factor use.  Individual factors are weighted in accordance with their respective prices, represented by the inner product $\boldsymbol{\pi}' \boldsymbol{f}$. The resulting optimization problem can therefore be expressed as a linear program of the form:
\begin{subequations}
\begin{align}
\min \quad & Z = \boldsymbol{\pi}' F^* x^* \\
\text{subject to} \quad & ({\cal I}^* - A^*) x^* \ge y
\end{align}
\end{subequations}
In this formulation:  
\begin{itemize}
    \item ${\cal I}^*$ ($n \times t$) denotes an incidence matrix, representing each sector's self-requirements across all alternative technologies.  
    \item $A^*$ ($n \times t$) is the augmented inter-industry transaction matrix, capturing the input requirements of each sector for every technology choice.  
    \item $x^*$ ($t \times 1$) is the vector of total outputs associated with each technology option.  
    \item $y$ ($t \times 1$) represents the final demand, remaining unchanged from the basic EIO model.  
    \item $F^*$ ($k \times t$) specifies factor requirement per unit of output.  
    \item $\boldsymbol{\pi}$ ($k \times 1$) is the vector of factor prices used to weight these intensities in the objective function.  
    \item and $\boldsymbol{f}$ ($k \times 1$) is the factor availability vector.
\end{itemize}
Together, these augmented matrices and vectors allow the RCOT model to systematically represent both sectoral interdependencies and technology-specific factor requirements, enabling the optimization of total factor use while accounting for alternative technological options.

\subsection{Model-Based Systems Engineering (MBSE)}\label{Sec:MBSEConcept}

In systems engineering, models serve as indispensable tools for understanding complex systems, enabling experiments, operations, and stakeholder negotiations by representing key aspects of the world from a specific perspective while simplifying it through abstractions that omit irrelevant details \cite{hmelo2006understanding,sterman1994learning,ramos2011model}. MBSE builds on this foundation as a standard, flexible modeling approach that uses a broad set of abstractions for the design and analysis of complex systems, including economic systems \cite{naderi2025convergent,Harris:2024:ISC-AP96}. Rather than treating economic, hydrologic, social, and other interlinked systems as isolated entities, MBSE provides a unified framework that integrates these domains, capturing both form (the physical or logical elements) and function (the processes that transform inputs into outputs) within a common modeling language \cite{de2022taxonomy,henderson2021value}.

In the context of economic modeling, MBSE act as a bridge between historically fragmented modeling approaches. For example, while economic models estimate changes in supply and demand, MBSE can facilitate their integration with land-use, hydrologic, ecosystem, energy, and other interconnected systems \cite{farid2022convergent}. This holistic perspective enables detailed graphical representations and simulations of economic processes within a common ontological framework and provides insight into their interactions within a larger SoS.
SysML provides a standardized set of diagrams for representing system architecture within MBSE. SysML diagrams exhibit three broad categories of systems thinking abstractions to graphically describe a system:  
\begin{enumerate*}
\item system boundary,
\item system form, and
\item system function  
\end{enumerate*}. This generic pattern constitutes a \emph{reference architecture} (RA)\cite{crawley2015system}, which can significantly reduce the cognitive complexity associated with understanding these systems.

Among the SysML diagrams, the Block Definition Diagram (BDD) captures the system form by specifying the hierarchy of its structural elements, their constituents and attributes, and how they are connected and/or related. In economic systems, these might represent physical elements like industrial sectors and production facilities. Relationships among these physical elements can be represented by a broad set of interconnection abstractions, such as encapsulation, decomposition, aggregation, generalization, and classification \cite{Delligatti:2014:00,naderi2025convergent}. Instantiation relationships are also important, as they describe connections at a generic level rather than specifying them for each and every instance \cite{holt2008sysml}. For example, in economic systems, the agricultural sector generally sells its products to specific industrial sectors.

Similarly, the Activity Diagram (ACT) describes what a system does and how it behaves \cite{Delligatti:2014:00,friedenthal2014practical}. ACT diagrams explicitly illustrate the flow of processes showing how activities transform inputs into outputs. For example, the transfer of energy from its source to the sectors that consume it to generate specific industrial products \cite{farid2022tensor,schoonenberg2019hetero}. In ACT diagrams, no restrictions are imposed on the type of system state; for instance, Lagrangian, power, and Hamiltonian variables may be defined over real numbers, complex numbers, integers, or even booleans. Moreover, continuous-time, discrete-time, discrete-event, and hybrid state evolutions can be straightforwardly simulated within ACT diagrams. These capabilities enable modelers to capture details of economic systems with enhanced accuracy, thereby reducing uncertainty and facilitating more rigorous analyses of complex interdependencies and dynamic behaviors. It should be noted that functional interactions in ACT and formal interfaces (i.e., associations) in BDD are equivalent only when the allocation of function to form is one-to-one \cite{crawley2015system,friedenthal2014practical}.

In BDD and ACT diagrams, the system boundary delineates between the system itself and everything else (in the system context)\cite{crawley2015system}. In SysML, the system boundary is always explicitly depicted by the diagram boundary, regardless of the diagram type. This boundary is labeled in the top left corner of each diagram with four pieces of information, presented in order:
\begin{enumerate*}
\item the type of diagram (e.g., BDD, ACT),
\item the type of model element (e.g., block, activity),
\item the name of the model element, and
\item the name of the diagram\cite{Delligatti:2014:00,naderi2025convergent}.
\end{enumerate*}
Whether a system boundary is open or closed is made explicit in SysML. If a block in BDD has a connecting arrow, it is considered an open system in its own right. Likewise, any activity in ACT with a connecting arrow is treated as an open system activity. In contrast, the lack of connecting arrows signifies a closed system. Explicitly defining the system boundary allows economic modelers to clearly distinguish between endogenous and exogenous factors, improving model structure, interpretation, and the reliability of scenario analyses.

Ultimately, SysML diagrams enhance shared understanding and support the co-production of knowledge when practitioners from diverse disciplines—including economics, hydrology, energy, and land use—conduct modeling within a common ontology \cite{moallemi2023knowledge}. This interdisciplinary coherence contributes to a modeling process that is more legitimate, credible, and salient \cite{cash2020salience,kunseler2015reflective}.

\subsection{Hetero-Functional Graph Theory (HFGT)}\label{Sec:HFGTConcept}

While MBSE, and specifically SysML, offers a unified modeling language \cite{little2023earth} for complex SoS, it inherently lacks the tools necessary for robust, large-scale quantitative analyses. Fortunately, HFGT \cite{Schoonenberg:2019:ISC-BK04,Farid:2022:ISC-J51,Farid:2016:ISC-BC06} provides an analytical framework for converting graphical SysML representations into formal mathematical and computational models. HFGT has been applied to single-domain systems such as electric power \cite{Farid:2015:SPG-J17,Thompson:2021:SPG-J46}, potable water \cite{Farid:2015:ISC-J19}, transportation \cite{Viswanath:2013:ETS-J08}, and mass-customized production systems \cite{Farid:2015:IEM-J23,Farid:2008:IEM-J05,Farid:2008:IEM-J04,Farid:2017:IEM-J13}. Its application extends to SoS, including multi-modal electrified transportation \cite{Farid:2016:ETS-J27,vanderWardt:2017:ETS-J33,Farid:2016:ETS-BC05}, microgrid-enabled production systems \cite{Schoonenberg:2017:IEM-J34}, hydrologic modeling \cite{harris2025integrative}, personalized healthcare delivery \cite{Khayal:2015:ISC-J20,Khayal:2017:ISC-J35,Khayal:2021:ISC-J48}, hydrogen-natural gas systems \cite{Schoonenberg:2022:ISC-J50}, the energy-water nexus \cite{Farid:2024:ISC-JR04}, and the American multi-modal energy system \cite{Thompson:2024:ISC-J55}. Notably, several of these SoS applications of HFGT integrate both time-driven and discrete-event dynamics.  

Additionally, recent studies have shown that MBSE-HFGT can formally generalize and extend process-based life-cycle assessment (LCA) \cite{gohil2025spatio, gohil2025enhancing}, whose conceptual foundations closely align with EIO systems. These studies have reconciled LCA with the shared language of MBSE-HFGT and have demonstrated that MBSE-HFGT is a formal generalization of process-based LCA. They also show how MBSE-HFGT can enhance the spatio-temporal resolution of LCA to align it with system design objectives \cite{gohil2025spatio} and adopt dynamic, data-driven approaches—such as real-time carbon intensity, operational adaptation, and cost fluctuations—to more accurately quantify environmental burdens. Together, these advances reflect the same underlying principles that support other extensions of EIO models—such as MRIO, EEIO, and ECEIO—which capture spatio-temporal dynamics of interdependent flows \cite{zhao2022global} to characterize system-wide sustainability outcomes.\\

\subsubsection{Essential Definitions}\label{subsec:HFGT}
Hetero-functional graphs (HFGs) are formal graphical models that represent the interconnectedness of complex SoS. HFGT is conceptually rooted in the universal structural principles of human language, particularly subjects and predicates, where predicates consist of verbs and objects \cite{Schoonenberg:2019:ISC-BK04,Farid:2022:ISC-J51}.  This reliance on the structure of human language provides the basis for a discipline-agnostic ontology that facilitates cross-disciplinary applications \cite{yang2019ontology}. 
In HFGT, a system is defined by a set of resources $R$, which act as the subjects; a set of processes $P$, which serve as predicates; and a set of operands $L$, which constitute the objects involved in these processes.   
\begin{defn}[System Operand 
\cite{SE-Handbook-Working-Group:2015:00}]
\label{Defn:D1}
An asset or object $l_i \in L$ that is operated on or utilized in the course of a process execution.
\end{defn}
\begin{defn}[System Process
\cite{Hoyle:1998:00,SE-Handbook-Working-Group:2015:00}]
\label{def:CH7:process}
An activity $p \in P$ that converts or transfers a specified set of input operands into a designated set of outputs. 
\end{defn}
\begin{defn}[System Resource
\cite{SE-Handbook-Working-Group:2015:00}]
\label{def:resources}
An asset or object $r_v \in R$ that enables the execution of a process.  
\end{defn}
The system resources $R=M \cup B \cup H$ are categorized into transformation resources $M$, independent buffers $B$, and transportation resources $H$.  The set of ``buffers" $B_S=M \cup B$ is introduced to support the discussion, and the system processes $P = P_\mu \cup P_{\bar{\eta}}$ are classified into transformation processes $P_\mu$ and refined transportation processes $P_\eta$.  This occurs from the concurrent execution of a transportation process and a holding process. HFGT further highlights that resources can perform one or more system processes, thereby generating a set of ``capabilities"\cite{Schoonenberg:2019:ISC-BK04}.
\begin{defn}[Buffer
\cite{Schoonenberg:2019:ISC-BK04,Farid:2022:ISC-J51}]
\label{defn:BSCh7}

A resource $r_v \in R$ is a buffer $b_s \in B_S$  if and only if it has the capacity to store or transform one or more operands at a specific spatial location.  
\end{defn}
\begin{defn}[Capability\cite{Schoonenberg:2019:ISC-BK04,Farid:2022:ISC-J51,Farid:2016:ISC-BC06}]
\label{defn:capabilityCh7}
An action $e_{wv} \in {\cal E}_S$ (in the SysML sense) defined by a system process $p_w \in P$ being executed by a resource $r_v \in R$.  It constitutes a subject + verb + operand sentence of the form: ``Resource $r_v$ does process $p_w$".  
\end{defn}

In HFGT, the engineering system meta-architecture must be instantiated and eventually translated into corresponding Petri net model(s) \cite{Farid:2022:ISC-J51}. To facilitate this, the positive and negative hetero-functional incidence tensors are introduced to characterize the flow of operands through buffers and capabilities.  
\begin{defn}[The Negative 3$^{rd}$ Order Hetero-functional Incidence Tensor (HFIT) $\widetilde{\cal M}_\rho^-$
\cite{Farid:2022:ISC-J51}]
\label{Defn:D6}
The negative hetero-functional incidence tensor $\widetilde{\cal M_\rho}^- \in \{0,1\}^{|L|\times |B_S| \times |{\cal E}_S|}$  is a third-order tensor whose element $\widetilde{\cal M}_\rho^{-}(i,y,\psi)=1$ when the system capability ${\epsilon}_\psi \in {\cal E}_S$ pulls operand $l_i \in L$ from buffer $b_{s_y} \in B_S$.
\end{defn} 

\begin{defn}[The Positive  3$^{rd}$ Order Hetero-functional Incidence Tensor (HFIT) $\widetilde{\cal M}_\rho^+$
\cite{Farid:2022:ISC-J51}]
The positive hetero-functional incidence tensor $\widetilde{\cal M}_\rho^+ \in \{0,1\}^{|L|\times |B_S| \times |{\cal E}_S|}$  is a third-order tensor whose element $\widetilde{\cal M}_\rho^{+}(i,y,\psi)=1$ when the system capability ${\epsilon}_\psi \in {\cal E}_S$ injects operand $l_i \in L$ into buffer $b_{s_y} \in B_S$.
\end{defn}
\noindent These incidence tensors are straightforwardly ``matricized" to form second-order Hetero-functional Incidence Matrices $M = M^+ - M^-$ with dimensions $|L||B_S|\times |{\cal E}|$. Consequently, the supply, demand, transportation, storage, transformation, assembly, and disassembly of multiple operands in distinct locations over time can be described by an Engineering System Net and its associated State Transition Function \cite{Schoonenberg:2022:ISC-J50}.

\begin{defn}[Engineering System Net
\cite{Schoonenberg:2022:ISC-J50}]
\label{Defn:ESN}
An elementary Petri net ${\cal N} = \{S, {\cal E}_S, \textbf{M}, W, Q\}$, where:
\begin{itemize}
    \item $S$ is the set of places with size: $|L||B_S|$,
    \item ${\cal E}_S$ is the set of transitions with size: $|{\cal E}|$,
    \item $\textbf{M}$ is the set of arcs, with the associated incidence matrices: $M = M^+ - M^-$,
    \item $W$ is the set of weights on the arcs, as captured in the incidence matrices,
    \item $Q=[Q_B; Q_E]$ is the marking vector for both the set of places and the set of transitions. 
\end{itemize}
\end{defn}

\begin{defn}[Engineering System Net State Transition Function
\cite{Schoonenberg:2022:ISC-J50}]
\label{Defn:ESN-STF}
The  state transition function of the engineering system net $\Phi()$ is:
\begin{equation}\label{CH6:eq:PhiCPN}
Q[k+1]=\Phi(Q[k],U^-[k], U^+[k]) \quad \forall k \in \{1, \dots, K\}
\end{equation}
where $k$ is the discrete time index, $K$ is the simulation horizon, $Q=[Q_{B}; Q_{\cal E}]$, $Q_B$ has size $|L||B_S| \times 1$, $Q_{\cal E}$ has size $|{\cal E}_S|\times 1$, the input firing vector $U^-[k]$ has size $|{\cal E}_S|\times 1$, and the output firing vector $U^+[k]$ has size $|{\cal E}_S|\times 1$.  
\begin{subequations}
\begin{align}\label{CH6:CH6:eq:Q_B:HFNMCFprogram}
Q_{B}[k+1]&=Q_{B}[k]+{M}^+U^+[k]\Delta T-{M}^-U^-[k]\Delta T \\ \label{CH6:CH6:eq:Q_E:HFNMCFprogram}
Q_{\cal E}[k+1]&=Q_{\cal E}[k]-U^+[k]\Delta T +U^-[k]\Delta T
\end{align}
\end{subequations}
where $\Delta T$ is the duration of the simulation time step.  
\end{defn}
\noindent It is crucial to mention that the engineering system net’s state transition function explicitly embodies continuity laws, allowing both the Eulerian and Lagrangian perspectives depending on the required modeling application.

In addition to the engineering system net, each operand in HFGT can have its own state and evolution.  This behavior is described by an Operand Net and its associated State Transition Function for each operand. 
\begin{defn}[Operand Net\cite{Farid:2008:IEM-J04,Schoonenberg:2019:ISC-BK04,Khayal:2017:ISC-J35,Schoonenberg:2017:IEM-J34}]\label{Defn:OperandNet} Given operand $l_i$, an elementary Petri net ${\cal N}_{l_i}= \{S_{l_i}, {\cal E}_{l_i}, \textbf{M}_{l_i}, W_{l_i}, Q_{l_i}\}$ where: 
\begin{itemize}
\item $S_{l_i}$ is the set of places describing the operand's state.  
\item ${\cal E}_{l_i}$ is the set of transitions describing the evolution of the operand's state.
\item $\textbf{M}_{l_i} \subseteq (S_{l_i} \times {\cal E}_{l_i}) \cup ({\cal E}_{l_i} \times S_{l_i})$ is the set of arcs, with the associated incidence matrices: $M_{l_i} = M^+_{l_i} - M^-_{l_i} \quad \forall l_i \in L$.  
\item $W_{l_i} : \textbf{M}_{l_i}$ is the set of weights on the arcs, as captured in the incidence matrices $M^+_{l_i},M^-_{l_i} \quad \forall l_i \in L$.  
\item $Q_{l_i}= [Q_{Sl_i}; Q_{{\cal E}l_i}]$ is the marking vector for both the set of places and the set of transitions. 
\end{itemize}
\end{defn}

\begin{defn}[Operand Net State Transition Function\cite{Farid:2008:IEM-J04,Schoonenberg:2019:ISC-BK04,Khayal:2017:ISC-J35,Schoonenberg:2017:IEM-J34}]\label{Defn:OperandNet-STF}
The  state transition function of each operand net $\Phi_{l_i}()$ is:
\begin{equation}\label{CH6:eq:PhiSPN}
Q_{l_i}[k+1]=\Phi_{l_i}(Q_{l_i}[k],U_{l_i}^-[k], U_{l_i}^+[k]) \quad \forall k \in \{1, \dots, K\} \quad i \in \{1, \dots, L\}
\end{equation}
where $Q_{l_i}=[Q_{Sl_i}; Q_{{\cal E} l_i}]$, $Q_{Sl_i}$ has size $|S_{l_i}| \times 1$, $Q_{{\cal E} l_i}$ has size $|{\cal E}_{l_i}| \times 1$, the input firing vector $U_{l_i}^-[k]$ has size $|{\cal E}_{l_i}|\times 1$, and the output firing vector $U^+[k]$ has size $|{\cal E}_{l_i}|\times 1$.  

\begin{subequations}
\begin{align}\label{X}
Q_{Sl_i}[k+1]&=Q_{Sl_i}[k]+{M_{l_i}}^+U_{l_i}^+[k]\Delta T - {M_{l_i}}^-U_{l_i}^-[k]\Delta T \\ \label{CH6:CH eq:Q_E:HFNMCFprogram}
Q_{{\cal E} l_i}[k+1]&=Q_{{\cal E} l_i}[k]-U_{l_i}^+[k]\Delta T +U_{l_i}^-[k]\Delta T
\end{align}
\end{subequations}
\end{defn}

\subsubsection{The Hetero-functional Network Minimum Cost Flow (HFNMCF) Problem}\label{subsec:HFNMCF}

HFGT simulates the behavior of an engineering system using the Hetero-Functional Network Minimum Cost Flow (HFNMCF) problem \cite{Schoonenberg:2022:ISC-J50}.  The HFNMCF problem extends the classical network minimum-cost flow problem to account for the heterogeneity of functions observed in Anthropocene SoS.  It optimizes the time-dependent flow and storage of multiple operands among buffers, enables transformation from one operand to another, and tracks the state of these operands.  In this regard, it is a highly flexible optimization problem applicable to a wide range of complex engineering systems.  For the purposes of this paper, the HFNMCF problem is a type of discrete-time-dependent, time-invariant, convex optimization program\cite{Schoonenberg:2022:ISC-J50}.

\begin{subequations}
\begin{align}\label{Eq:ObjFunc}
Z &= \sum_{k=1}^{K-1} x^T[k] F_{QP} x[k] + f_{QP}^T x[k]
\end{align}
\begin{align}\label{Eq:ESN-STF1}
\text{s.t. } -Q_{B}[k+1]+Q_{B}[k]+{M}^+U^+[k]\Delta T - {M}^-U^-[k]\Delta T=&0 && \!\!\!\!\!\!\!\!\!\!\!\!\!\!\!\!\!\!\!\!\!\!\!\!\!\!\!\!\!\!\!\!\!\!\!\!\!\!\!\!\!\forall k \in \{1, \dots, K\}\\  \label{Eq:ESN-STF2}
-Q_{\cal E}[k+1]+Q_{\cal E}[k]-U^+[k]\Delta T + U^-[k]\Delta T=&0 && \!\!\!\!\!\!\!\!\!\!\!\!\!\!\!\!\!\!\!\!\!\!\!\!\!\!\!\!\!\!\!\!\!\!\!\!\!\!\!\!\!\forall k \in \{1, \dots, K\}\\ \label{Eq:DurationConstraint}
 - U_\psi^+[k+k_{d\psi}]+ U_{\psi}^-[k] = &0 && \!\!\!\!\!\!\!\!\!\!\!\!\!\!\!\!\!\!\!\!\!\!\!\!\!\!\!\!\!\!\!\!\!\!\!\!\!\!\!\!\!\forall k\in \{1, \dots, K\} \quad \psi \in \{1, \dots, {\cal E}_S\}\\ \label{Eq:OperandNet-STF1}-Q_{Sl_i}[k+1]+Q_{Sl_i}[k]+{M}_{l_i}^+U_{l_i}^+[k]\Delta T - {M}_{l_i}^-U_{l_i}^-[k]\Delta T=&0 && \!\!\!\!\!\!\!\!\!\!\!\!\!\!\!\!\!\!\!\!\!\!\!\!\!\!\!\!\!\!\!\!\!\!\!\!\!\!\!\!\!\forall k \in \{1, \dots, K\} \quad i \in \{1, \dots, |L|\}\\ \label{Eq:OperandNet-STF2}
-Q_{{\cal E}l_i}[k+1]+Q_{{\cal E}l_i}[k]-U_{l_i}^+[k]\Delta T + U_{l_i}^-[k]\Delta T=&0 && \!\!\!\!\!\!\!\!\!\!\!\!\!\!\!\!\!\!\!\!\!\!\!\!\!\!\!\!\!\!\!\!\!\!\!\!\!\!\!\!\!\forall k \in \{1, \dots, K\} \quad i \in \{1, \dots, |L|\}\\ \label{Eq:OperandNetDurationConstraint}
- U_{xl_i}^+[k+k_{dxl_i}]+ U_{xl_i}^-[k] = &0 &&  \!\!\!\!\!\!\!\!\!\!\!\!\!\!\!\!\!\!\!\!\!\!\!\!\!\!\!\!\!\!\!\!\!\!\!\!\!\!\!\!\!
\forall k\in \{1, \dots, K\}, \: \forall x\in \{1, \dots, |{\cal E}_{l_i}\}|, \: l_i \in \{1, \dots, |L|\}\\ \label{Eq:SyncPlus}
U^+_L[k] - \widehat{\Lambda}^+ U^+[k] =&0 && \!\!\!\!\!\!\!\!\!\!\!\!\!\!\!\!\!\!\!\!\!\!\!\!\!\!\!\!\!\!\!\!\!\!\!\!\!\!\!\!\!\forall k \in \{1, \dots, K\}\\ \label{Eq:SyncMinus}
U^-_L[k] - \widehat{\Lambda}^- U^-[k] =&0 && \!\!\!\!\!\!\!\!\!\!\!\!\!\!\!\!\!\!\!\!\!\!\!\!\!\!\!\!\!\!\!\!\!\!\!\!\!\!\!\!\!\forall k \in \{1, \dots, K\}\\ \label{CH6:eq:HFGTprog:comp:Bound}
\begin{bmatrix}
D_{Up} & \mathbf{0} \\ \mathbf{0} & D_{Un}
\end{bmatrix} \begin{bmatrix}
U^+ \\ U^-
\end{bmatrix}[k] =& \begin{bmatrix}
C_{Up} \\ C_{Un}
\end{bmatrix}[k] && \!\!\!\!\!\!\!\!\!\!\!\!\!\!\!\!\!\!\!\!\!\!\!\!\!\!\!\!\!\!\!\!\!\!\!\!\!\!\!\!\!\forall k \in \{1, \dots, K\} \\\label{Eq:OperandRequirements}
\begin{bmatrix}
E_{Lp} & \mathbf{0} \\ \mathbf{0} & E_{Ln}
\end{bmatrix} \begin{bmatrix}
U^+_{l_i} \\ U^-_{l_i}
\end{bmatrix}[k] =& \begin{bmatrix}
F_{Lpi} \\ F_{Lni}
\end{bmatrix}[k] && \!\!\!\!\!\!\!\!\!\!\!\!\!\!\!\!\!\!\!\!\!\!\!\!\!\!\!\!\!\!\!\!\!\!\!\!\!\!\!\!\!\forall k \in \{1, \dots, K\}\quad i \in \{1, \dots, |L|\} \\\label{CH6:eq:HFGTprog:comp:Init} 
\begin{bmatrix} Q_B ; Q_{\cal E} ; Q_{SL} \end{bmatrix}[1] =& \begin{bmatrix} C_{B1} ; C_{{\cal E}1} ; C_{{SL}1} \end{bmatrix} \\ \label{CH6:eq:HFGTprog:comp:Fini}
\begin{bmatrix} Q_B ; Q_{\cal E} ; Q_{SL} ; U^- ; U_L^- \end{bmatrix}[K+1] =   &\begin{bmatrix} C_{BK} ; C_{{\cal E}K} ; C_{{SL}K} ; \mathbf{0} ; \mathbf{0} \end{bmatrix}\\ \label{ch6:eq:QPcanonicalform:3}
\underline{E}_{CP} \leq D(X) \leq& \overline{E}_{CP} \\ \label{Eq:DeviceModels}
g(X,Y) =& 0 \\ \label{Eq:DeviceModels2}
h(Y) \leq& 0 
\end{align}
\end{subequations}
where $X=\left[x[1]; \ldots; x[K]\right]$  is the vector of primary decision variables, and $Y=\left[y[1]; \ldots; y[K]\right]$  is the vector of auxiliary decision variables at time $k$.

\section{A Motivational Example: Applying MBSE-HFGT to an EIO model}
\label{Sec:Methodology}

To support the remainder of the paper's contribution, this section introduces a motivational example to compare and contrast EIO models with MBSE and HFGT. Subsection \ref{subsec:duchinexample} describes the RCOT model example. Subsection \ref{referencearch_EIO} presents the development of the economic system RA under the HFGT meta-architecture to represent the form and the function of the RCOT example. Finally, Subsection \ref{Sec:HFGT} derives a system of equations in accordance with the HFNMCF problem outlined in Subsection \ref{subsec:HFNMCF} to quantitatively show how MBSE-HFGT reproduces the same results as the traditional RCOT model. This provides an instructive foundation, illustrating the methodology in a scenario that is both tractable and conceptually rich.

\subsection{Economic Input Output (EIO) Model} \label{subsec:duchinexample}

\liinesbigfig{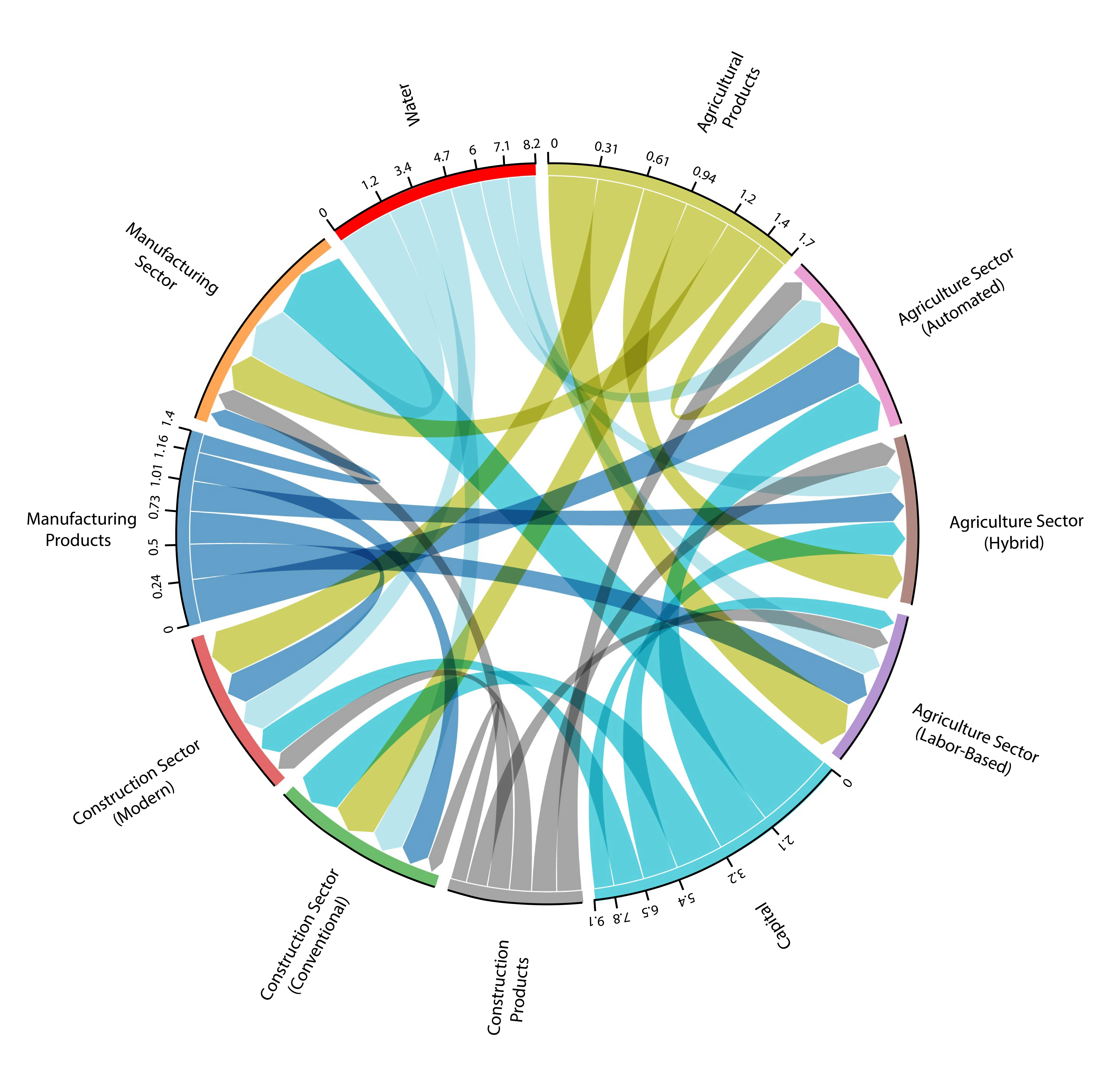}{Inter-industrial relationships of economic sectors in terms of technical coefficients }{fig:Chord}{5.5}

This section describes an example economic system using the Rectangular Choice of Technology (RCOT) model described by Duchin \cite{duchin2011sectors}. The synthetic economy example is illustrated in Fig \ref{fig:Chord}.  It is specified by the following matrices and vectors that constitute the RCOT formulation.

\[
{\cal I^*} = 
\begin{bmatrix}
1 & 0 & 0 & 0 & 0 & 0 \\
0 & 1 & 1 & 0 & 0 & 0 \\
0 & 0 & 0 & 1 & 1 & 1
\end{bmatrix}_{3 \times 6}, 
\quad
A^* = 
\begin{bmatrix}
0.35 & 0.15 & 0.23 & 0.26 & 0.28 & 0.24 \\
0.25 & 0.22 & 0.16 & 0.22 & 0.21 & 0.25 \\
0.20 & 0.26 & 0.30 & 0.31 & 0.33 & 0.30
\end{bmatrix}_{3 \times 6}
\]

\[
F^* = 
\begin{bmatrix}
2.1 & 3.2 & 1.9 & 1.2 & 0.8 & 1.4 \\
1.2 & 2.2 & 1.3 & 1.3 & 1.1 & 1.1
\end{bmatrix}_{2 \times 6}, 
\quad
y = 
\begin{bmatrix}
20 \\ 25 \\ 22
\end{bmatrix}_{3 \times 1},
\quad
\quad
f = 
\begin{bmatrix}
540 \\ 342
\end{bmatrix}_{2 \times 1},
\quad
\pi =
\begin{bmatrix}
1 \\ 0.9
\end{bmatrix}_{2 \times 1}.
\]

As quantified in the matrices above, the economic system comprises $n=3$ sectors and $k=2$ primary factors of production. The sectors differ in the number of technological alternatives available to them. For clarity, technological options are conventionally assigned to the following sectors.  Sector 1 (manufacturing) is limited to a single technology ($t_1=1$).  Sector 2 (construction) has access to two alternative technologies ($t_2=2$).  Sector 3 (agriculture) can operate with three technological options ($t_3=3$).  This configuration yields a total of $ T=6$ distinct technologies for the economy as a whole. The factors of production are restricted to two essential inputs:  water and capital.  Together, they span the system's resource requirements.

Solving the RCOT optimization problem yields the optimal value Z, the final output vector, $\mathbf{x}$, and the corresponding factor use vector, $\boldsymbol{\phi}$.  
\[
Z = 805.724,
\quad
\mathbf{x} =
\begin{bmatrix}
99.7883 \\ 0 \\ 87.5364 \\ 0 \\ 26.6400 \\ 71.9500
\end{bmatrix}_{6 \times 1},
\quad
\boldsymbol{\phi} =
\begin{bmatrix}
498.92 \\ 342.00
\end{bmatrix}_{2 \times 1}.
\]

\subsection{Model-Based Systems Engineering (MBSE)}
\label{referencearch_EIO}

Next, the EIO example introduced in the previous subsection is modeled as a SysML RA (as described previously in Sec. \ref{Sec:MBSEConcept}). By situating the EIO model within a SysML RA, the form and function of the economic system's architecture is elucidated.  To that end, a domain-specific economic system RA is developed as a special case of the (domain-independent) HFGT meta-architecture.  Fig. \ref{fig:EIO_BDD} shows the form of the economic system as a BDD.  Importantly, core HFGT concepts — such as resources, processes, and operands — are mapped to their counterparts in the EIO model. Fig. \ref{fig:EIO_ACT} shows the function of the economic system as an ACT.  It graphically depicts intersectoral transactions, factor-use allocations, and final-demand flows.  It is important to recognize that the reference architecture of the economic system can be readily extended, either within the ACT by adding new functions (e.g., economic activities) or within the BDD by incorporating new regional economies or entirely different (non-economic) systems.  

\subsection{Hetero-functional Graph Theory (HFGT)}\label{Sec:HFGT}
The domain-independent HFGT definitions introduced in Section~\ref{Sec:HFGTConcept} acquire domain-specific significance in the context of the EIO system model introduced in Sec. \ref{subsec:duchinexample}.  More specifically, the system operands $L$ (Def. \ref{Defn:D1}) are water, capital, and the outputs from the manufacturing, construction, and agricultural sectors. The system processes $P$ (Def.~\ref{def:CH7:process}) appear as the six operations in ``The Economy" block in Fig. ~\ref{fig:EIO_BDD}.  Collectively, they correspond to the economic mechanisms responsible for the production of agricultural, manufacturing, and construction products.  The system resources $R$ (Def. \ref{def:resources}) carry out the system processes.  In the context of the EIO model, the whole economy is a transformation resource $(M)$ that realizes all the processes. Because the EIO model focuses on functions, their inputs, and their outputs, there are no other system resources.  The capabilities ${\cal E}_S$ (Def. \ref{defn:capabilityCh7}) combined the resources and processes into subject + predicate sentences.  For example, the economy produces agricultural products with automated technologies. 

\liinesbigfig{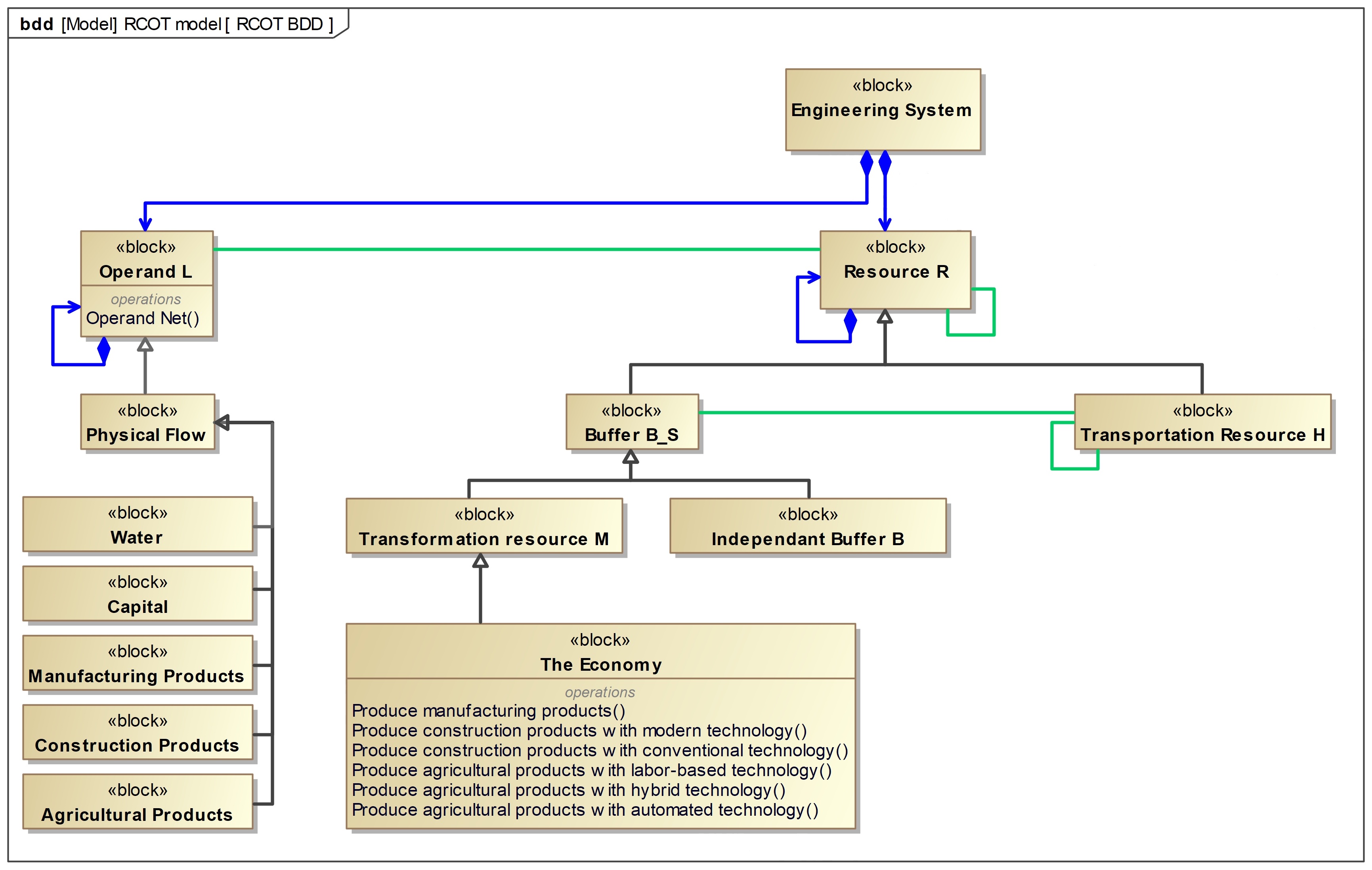}{Block Definition Diagram that defines the EIO model form.}{fig:EIO_BDD}{5}
\liinesbigfig{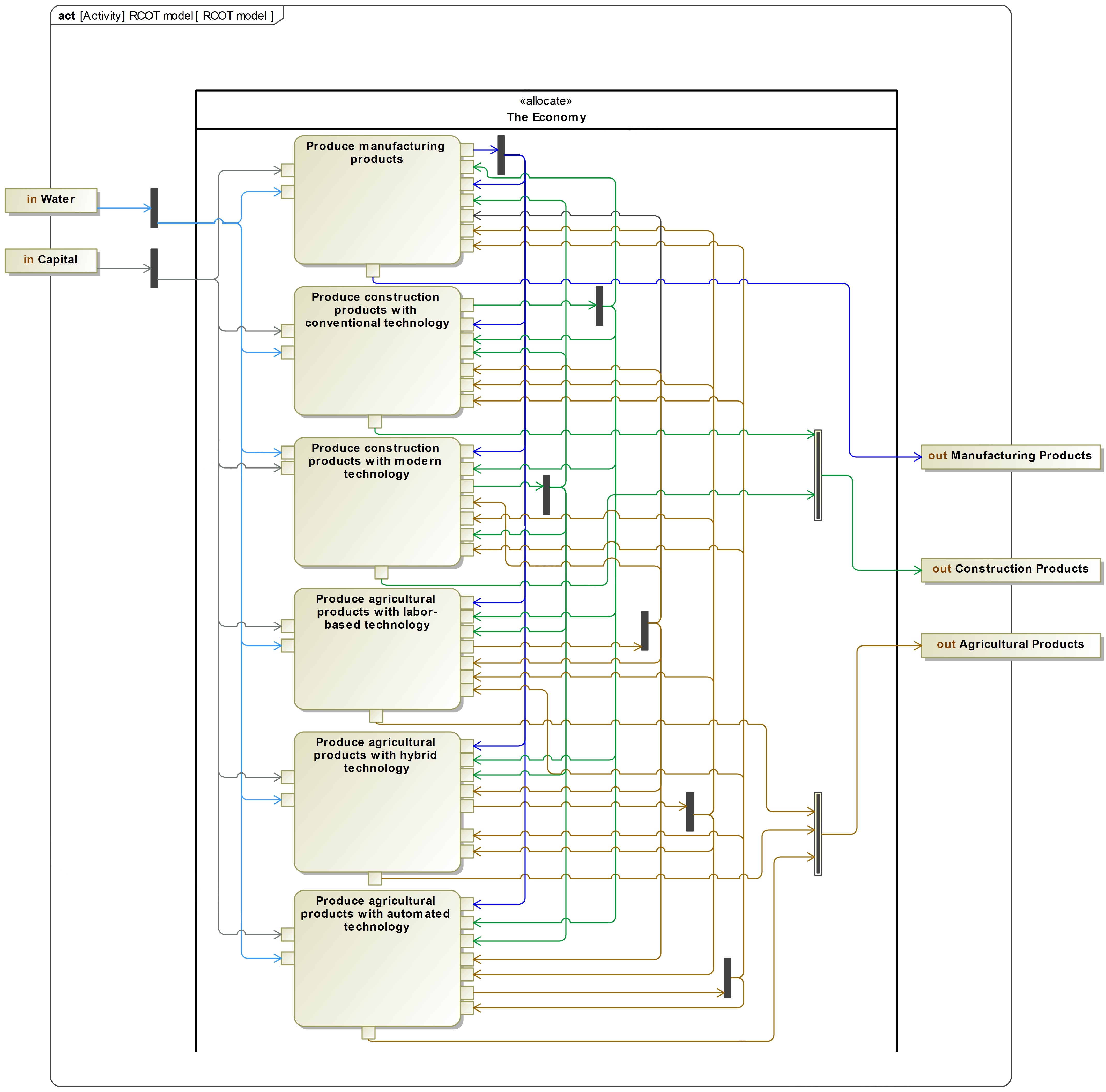}{Activity Diagram that defines the EIO model function.}{fig:EIO_ACT}{5}

To internalize the objective function and the constraint of the RCOT model (i.e., Eqs.~\ref{firstconseio} and \ref{secondconseio}) within HFGT, we reformulate them as
the following equation:  
\begin{equation}
\Big[ \frac{{\cal I^*} - A^*}{-F} \Big] x \ge \Big[ \frac{y}{-f} \Big]
\end{equation}

The right-hand side represents a vertically concatenated vector composed of the final demand ($y$) and factor availability ($f$), given by:

\begin{equation}
\left[\frac{y}{-f}\right] =
\begin{bmatrix}
\text{Manufacturing products} \\
\text{Construction products} \\
\text{Agricultural products} \\
\text{Capital} \\
\text{Water}
\end{bmatrix}
\label{alloperands}
\end{equation}



The final output vector $x$ is also expressed as:

\begin{equation}
x =
\begin{bmatrix}
\text{Economy produces manufactured products} \\
\text{Economy produces construction products with conventional technology} \\
\text{Economy produces construction products with modern technology} \\
\text{Economy produces agricultural products with labor-based technology} \\
\text{Economy produces agricultural products with hybrid technology} \\
\text{Economy produces agricultural products with automated technology}
\end{bmatrix}
\end{equation}

Each element of the vector $x$ represents a distinct capability associated with the economic system, characterizing a specific process undertaken to produce inter-industry outputs (see Fig. \ref{fig:EIO_BDD} and \ref{fig:EIO_ACT}). In other words, each entry in this vector adheres to the “Subject + Verb + Object” construct of the HFGT framework described in Sec.~\ref{Sec:HFGTConcept}.

The configuration of the engineering system net, represented by $Q_B$, is designed to capture the heterogeneity of system buffers and operands. Accordingly, $Q_B$ is organized as a composite structure consisting of five vertically concatenated vectors:

\begin{align}
Q_B = \left[ 
Q_{\text{Economy-ManProd}},\; 
Q_{\text{Economy-ConsProd}},\; 
Q_{\text{Economy-AgProd}},\; 
Q_{\text{Economy-Water}},\; 
Q_{\text{Economy-Capital}}
\right]
\end{align}
The notation $ Q_{\text{Economy}} $ is introduced to represent the five elements of $ Q_B $ corresponding to the set of five operands, expressed in million-dollar units except for water, which is given in million-gallon units.
Subsequently, the transitions of the engineering system network, $U$, are structured to capture the distinct capabilities of the economic system. Accordingly, $U$ is organized as a composite entity comprising six vertically concatenated vectors.

\begin{align*}
U = \big[ 
& U_{\text{ProduceManProd}},\;
  U_{\text{ProduseConsProd-ConvTech}},\;
  U_{\text{ProduseConsProd-ModTech}}, \\
& U_{\text{ProduceAgProd-LaborTech}},\;
  U_{\text{ProduceAgProd-HybridTech}},\;
  U_{\text{ProduceAgProd-AutoTech}} 
\big]
\end{align*}

Owing to the inherent structural regularities and simplifying characteristics of the EIO model, the general formulation of the HFNMCF optimization problem presented in Eqs.~\ref{Eq:ObjFunc} to \ref{Eq:DeviceModels2} reduces to the following specialized form:

\begin{align}
\text{minimize} \; Z &= \boldsymbol{\pi}' F^* U
\label{objHFGT} \\
\label{constraintHFGT}
M\,U &\ge C
\end{align}

\begin{itemize}
    \item The objective function in Eq.~\ref{Eq:ObjFunc} simplifies to Eq.~\ref{objHFGT}, conveying the same formulation, with the distinction that, as discussed above, $x$=$U$ in this instance.
\item Eq.~\ref{Eq:ESN-STF1} simplifies to Eq.~\ref{constraintHFGT}, where $ C = \Big[\frac{y}{-f}\Big] $ represents the vector encompassing all operands within the economic system.  
\end{itemize}
The equality constraint in Eq. \ref{Eq:ESN-STF1} was reformulated as the inequality in Eq. \ref{constraintHFGT} because the basic EIO formulation omits the marking vector $Q$ at both the current and future time steps.  In effect, if the inequality in the EIO model were converted into an equality with its associated surplus variable\cite{boyd2004convex,freund1991polynomial,picheny2016bayesian}, it would represent the surplus found in the marking vector $Q$.  The elimination of storage in a surplus variable removes the need for temporal modeling in the EIO model.  

The original rationale for using an inequality in EIO formulations is not clearly documented. However, this paper hypothesizes that this relaxation was introduced to accommodate empirical data limitations. Historically, EIO tables were compiled from surveys and self-reported statistics, which often contained measurement errors, rounding inconsistencies, or incomplete entries. By expressing flow balances as inequalities rather than strict equalities, modelers could obtain feasible solutions even when the reported data violated perfect conservation, thereby tolerating uncertainty and structural inconsistencies \cite{palamodov1970linear}.

To demonstrate the methodology introduced in \ref{Sec:Methodology}, this section instantiates a canonical EIO configuration, described in Subsection \ref{subsec:duchinexample}. To decipher the mechanism of the engineering system net incidence matrix, the system is first represented as a colored Petri net in Fig \ref{fig:PetriNet}. This Petri net is the instantiated version of the example system described in Subsection \ref{subsec:duchinexample} within the economic system RA (Sec. \ref{referencearch_EIO}) which captures the dynamics of operands (i.e. both factors of production and inter-industry products) as they flow through the economic system. In the Petri net, operands are contained within discrete places (circles). Water (blue), capital (gray), agricultural products (green), manufacturing products (brown), and construction products (orange) are represented as tokens that undergo transformation processes through transitions—equivalent to the system’s capabilities—which encode the production and exchange mechanisms governing the flow of operands within the system.

\liinesbigfig{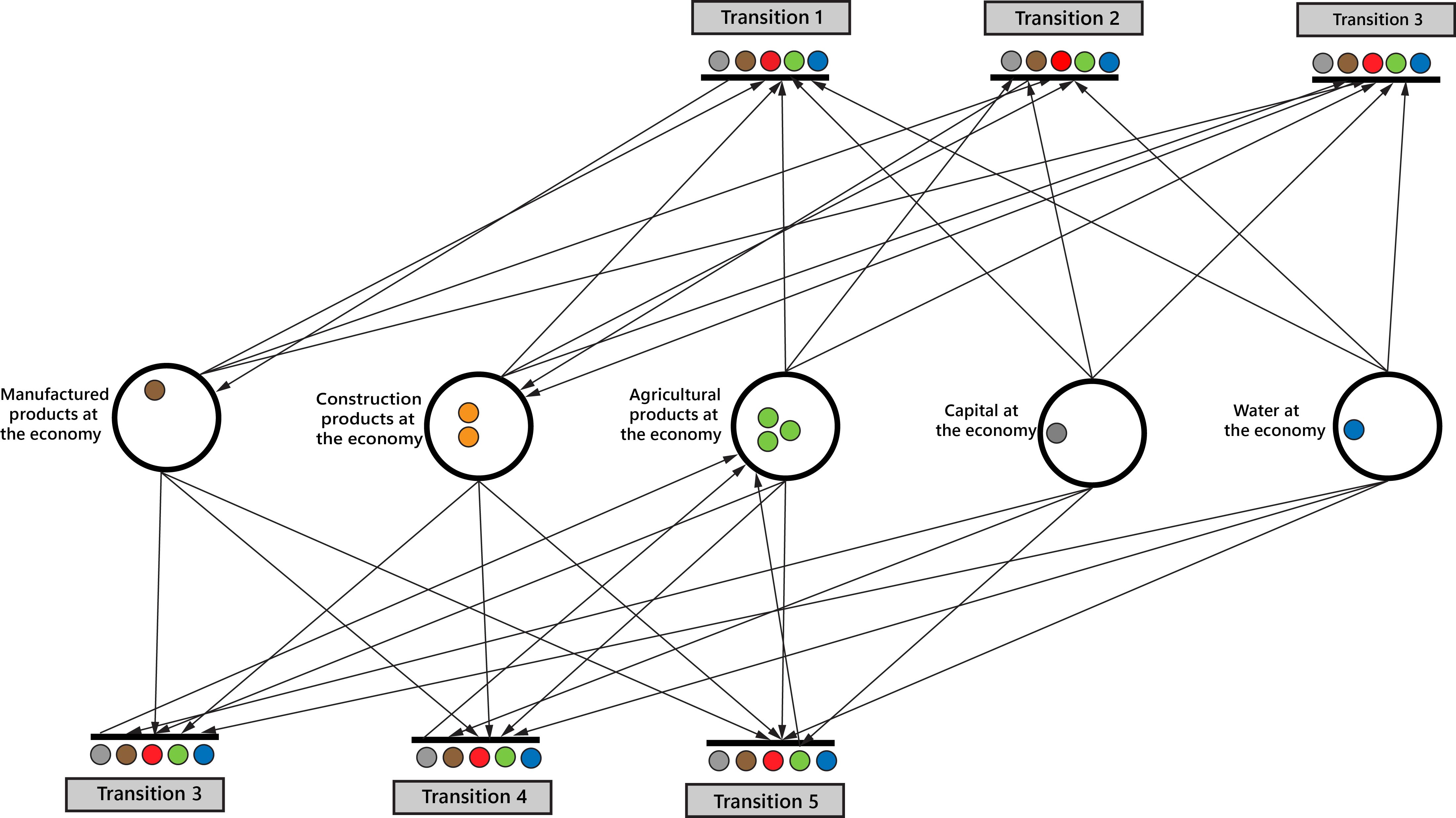}{Colored Petri Net representation of the EIO model}{fig:PetriNet}{6}

The Petri net, along with its underlying architectural schema, is encoded in an XML file and processed using the HFGT toolbox \cite{Thompson:2020:00}. This toolbox generates the hetero-functional incidence tensors and their associated metadata in JSON format.  Therefore, the engineering system net incidence matrix takes a block matrix form:

\begin{align*}
M^{+} &=
\begin{bmatrix}
1 & 0 & 0 & 0 & 0 & 0 \\
0 & 1 & 1 & 0 & 0 & 0\\
0 & 0 & 0 & 1 & 1 & 1\\
0 & 0 & 0 & 0 & 0 & 0\\
0 & 0 & 0 & 0 & 0 & 0
\end{bmatrix},
\quad
M^{-} =
\begin{bmatrix}
0.35 & 0.15 & 0.23 & 0.26 & 0.28 & 0.24 \\
0.25 & 0.22 & 0.16 & 0.22 & 0.21 & 0.25\\
0.20 & 0.26 & 0.30  & 0.31 & 0.33 & 0.30\\
2.1 & 3.2 & 1.9 & 1.2 & 0.8 & 1.4\\
1.2 & 2.2 & 1.3 & 1.3 & 1.1 & 1.1
\end{bmatrix},
\\[1ex]
M &= M^{+}-M^{-} =
\begin{bmatrix}
0.65 & -0.15 & -0.23 & -0.26 & -0.28 & -0.24 \\
-0.25 & 0.78 & 0.84 & -0.16 & -0.22 & -0.25\\
-0.20 & -0.26 & -0.30 & 0.69 & 0.67 & 0.70\\
-2.1 & -3.2 & -1.9 & -1.2 & -0.8 & -1.4\\
-1.2 & -2.2 & -1.3 & -1.3 & -1.1 & -1.1
\end{bmatrix}
\end{align*}

The JSON output from the HFGT toolbox is then imported into a Julia-based simulation environment where the HFNMCF problem is solved. The simulation results for the synthetic EIO model show complete agreement between the RCOT model and the HFNMCF problem (Table \ref{HFGTresults}). The solution of the HFNMCF problem demonstrates how the MBSE-HFGT framework can capture the dynamics of an RCOT model. More importantly, these results indicate that the RCOT model can be regarded as a special case of the MBSE-HFGT framework, suggesting that more complex versions of the EIO model—such as MRIO and EEIO—can also be represented within the MBSE-HFGT framework.

\begin{table}[ht]
\centering
\caption{Optimized values for capabilities ($U$) (i.e., final output ($x$) in RCOT), objective value ($Z$), and factor use ($\boldsymbol{\phi}^*$).}
\label{tab:opt_results}
\begin{tabular}{llr}
\toprule
 Capability & Value & Percentage of total\\
\midrule
 Economy produces manufactured products 
   & 99.7883 & 33.0\% \\
 Economy produces construction products with conventional technology 
   & 0.0000  & 0.0\% \\
 Economy produces construction products with modern technology 
   & 87.5364 & 29.0\% \\
 Economy produces agricultural products with labor-based technology 
   & 0.0000  & 0.0\% \\
 Economy produces agricultural products with hybrid technology 
   & 26.6439 & 8.8\% \\
 Economy produces agricultural products with automated technology  
   & 71.9531 & 23.8\% \\
\midrule
  & \textbf{Total}       & \textbf{286} \\
  & \textbf{Objective $Z$} & 805.7241 \\
\midrule
\multicolumn{3}{l}{\textbf{Factor use $\boldsymbol{\phi}^*$}} \\
\multicolumn{1}{l}{Capital} & \multicolumn{2}{r}{498.92 (Million \$)} \\
\multicolumn{1}{l}{Water}   & \multicolumn{2}{r}{342.00 (Million gallons)} \\
\bottomrule
\label{HFGTresults}
\end{tabular}
\end{table}

\section{Discussion}\label{Sec:DiscussionandResults}
Internalizing EIO models within the MBSE–HFGT framework demonstrates the scalability, extensibility, and integrative capacity of this framework for modeling economic systems. The RCOT synthetic example served as a pedagogical foundation, confirming that modeling multi-domain economic systems with detailed representation of heterogeneous system entities, variables, interdependencies, and structural dynamics can be addressed using the broad set of abstractions and a common language in SysML's rich graphical ontology. MBSE expressed in SysML provides a nuanced means of defining system boundaries, form, and function, while capturing the inherent nature of diverse entities and their interdependencies within EIO models.

A key novelty of this work lies in the graphical representation of the economic system’s structural form and function within SysML, presented here for the first time. Moreover, the RCOT example confirms that the core inter-industrial dynamics of an EIO model can be represented through the unified mathematical structure of HFGT, reproducing results equivalent to those obtained from the standard EIO approach. 
Extending the approach to other economic models with considerably greater complexity that are rooted in EIO fundamentals should be achievable through the framework’s extensibility. In such cases, system-specific tensors are automatically generated via the HFGT toolbox, enabling more complex dynamics to be incorporated without altering the underlying ontology.

Together, these results highlight the broader value of MBSE–HFGT in supporting the simulation of complex, heterogeneous EIO systems. Such heterogeneity includes algebraic, differential, and differential-algebraic equations, whether linear or nonlinear, as well as deterministic and stochastic models with continuous and discrete variables. In this unified modeling framework, the many features of SysML and HFGT provide an effective computational approach capable of handling the diverse characteristics of Anthropocene SoS, including economic systems. Representing all models within a common ontology also enables knowledge co-production, offering a pathway to address the complex challenges of contemporary sustainability more effectively than traditional approaches\cite{norstrom2020principles,moallemi2020exploratory}. Applying this comprehensive approach aligns with the National Science Foundation’s Growing Convergence Research (GCR) initiative, which aims for deep integration across disciplines to address pressing societal challenges \cite{nsf-gcr}.

\section{ Conclusion and future work}\label{Sec:Conclusion}

To integratively address the interdependent societal challenges of the Anthropocene, an SoS convergence paradigm that includes a computational framework, decision-support system, and educational pedagogy is required \cite{little2023earth}. However, the vast number of interacting systems and the heterogeneity of their discipline-specific ontologies make the direct integration of their models practically infeasible. The MBSE–HFGT framework provides researchers and practitioners across diverse domains—such as hydrology, land use, economics, energy, and healthcare—with a pragmatic and systematic tool for integratively modeling and addressing complex problems that propagate across interconnected systems with a unified modeling language.

This study presents the first application of the MBSE–HFGT framework to internalize EIO models within a consistent ontological structure and demonstrates that it can be treated as an interconnected system within a broader complex SoS. By facilitating cross-domain representation and integration, this framework also supports progress toward achieving the diverse United Nations Sustainable Development Goals (SDGs) \cite{UN2025}.

The SoS convergence paradigm is currently being implemented and validated for three interdependent societal challenges—eutrophication, agricultural impacts, and economic growth—within the Chesapeake Bay Watershed. The Chesapeake Bay Program (CBP) operates a sophisticated watershed management system that balances competing priorities across multiple federal, state, and private institutions. Building on decades of CBP research that has advanced modeling of land-use, watershed and estuarine systems \cite{hood2021chesapeake}, this work extends the existing framework by incorporating a model of an economic system.

\section*{Acknowledgments}\label{Sec:Acknowledgments}
This research is based on work supported by the Growing Convergence Research Program of the National Science Foundation under Grant Numbers OIA 2317874 and OIA 2317877. 
\section*
{Software and Data Availability }
All the codes and data used in this research are publicly available at the GitHub repository:
https://github.com/michaelnaderi/MonoLake.git
\section*
{CRediT author statement}
\textbf{Mohammad Mahdi Naderi}: Conceptualization, Data curation, Investigation, Methodology, Resources, Software, Validation, Visualization, Writing – original draft, Writing – review \& editing, Formal analysis.  
\textbf{Megan S. Harris}: Methodology, Software, Validation, Writing – review \& editing.   
\textbf{John C. Little}: Methodology, Funding acquisition, Supervision, Writing – review \& editing.  
\textbf{Amro M. Farid}: Conceptualization, Funding acquisition, Supervision, Validation, Writing – review \& editing.

\bibliographystyle{IEEEtran}
\bibliography{LIINESLibrary,LIINESPublications,references}

\end{document}